# Synchrotron radiation microtomography of brain hemisphere and spinal cord of a mouse model of multiple sclerosis revealed a correlation between capillary dilation and clinical score

**Running title:** Vasodilation and vacuolation in MS model

Rino Saiga[1], Masato Hoshino[2], Akihisa Takeuchi[2], Kentaro Uesugi[2], Katsuko Naitou[3], Akemi Kamijo[3], Noboru Kawabe[3], Masato Ohtsuka[4], Shunya Takizawa[5], and Ryuta Mizutani[1,*]

[1]Department of Applied Biochemistry, Tokai University, Hiratsuka, Kanagawa 259-1292, Japan
[2]Japan Synchrotron Radiation Research Institute (JASRI/SPring-8), Sayo, Hyogo 679-5198, Japan
[3]Support Center for Medical Research and Education, Tokai University, Isehara, Kanagawa 259-1193, Japan
[4]Department of Molecular Life Science, Division of Basic Medical Science and Molecular Medicine, Tokai University School of Medicine, Isehara, Kanagawa 259-1193, Japan
[5]Department of Neurology, Tokai University School of Medicine, Isehara, Kanagawa 259-1193, Japan
*Correspondence: mizutanilaboratory@gmail.com

**Abstract**
Multiple sclerosis is a neurological disorder in which the myelin sheaths of axons are damaged by the immune response. We report here a three-dimensional structural analysis of brain and spinal cord tissues of a mouse model of multiple sclerosis, known as experimental autoimmune encephalomyelitis (EAE). EAE-induced mice were raised with or without administration of fingolimod, which is used in the treatment of multiple sclerosis. Brains and spinal cords dissected from the EAE mice were lyophilized so as to reconstitute the intrinsic contrast of tissue elements, such as axons, in X-ray images. Three-dimensional structures of the brain hemispheres and spinal cords of the EAE mice were visualized with synchrotron radiation microtomography. Microtomographic cross sections reconstructed from the X-ray images revealed dilation of capillary vessels and vacuolation in the spinal cord of the EAE mice. Vacuolation was also observed in the cerebellum, suggesting that the neuroinflammatory response progressed in the brain. The vessel networks and vacuolation lesions in the spinal cords were modelled by automatically tracing the three-dimensional image in order to analyze the tissue structures quantitatively. The results of the analysis indicated that the distribution of vacuolations was not uniform but three-dimensionally localized. The mean vessel diameter showed a linear correlation with the clinical score, indicating that vasodilation is relevant to paralysis severity in the disease model. We suggest that vasodilation and vacuolation are related with neurological symptoms of multiple sclerosis.

**Keywords:** synchrotron radiation, x-ray microtomography, multiple sclerosis, experimental autoimmune encephalomyelitis, vasodilation, vacuolation

**1. Introduction**

Multiple sclerosis (MS) is a neurological disorder in which myelinated axons of the central nervous system are damaged by the immune response (Compston & Coles, 2008). It has been suggested that axonal loss in brain and spinal cord tissue (Suzuki et al., 1969; Bitsch et al., 2000; DeLuca et al., 2006) correlates with irreversible neurological impairment (Bjartmar et al., 2003). It has also been shown that axonal damage is associated with inflammation in brain tissue (Frischer et al., 2009). The disease process is considered to be initiated by autoreactive lymphocytes that cause inflammatory responses in the central nervous system (Dendrou et al., 2015).

Experimental autoimmune encephalomyelitis (EAE) is an animal model of MS (Constantinescu et al., 2011; Robinson et al., 2014). A number of elements of the neuroinflammatory responses observed in MS have been identified in EAE-induced animal models (Gold et al., 2006; Steinman and Zamvil, 2006; Farooqi et al., 2010). Although EAE is a reductive model (Dendrou et al., 2015) in which the autoimmune response is artificially induced using adjuvants, EAE has widely been used for studying pathogenesis of MS as well as potential therapeutic interventions (Robinson et al., 2014). Since EAE is characterized by neuroinflammation (Brown & Sawchenko, 2007), EAE-induced animal models have been used for analyzing the efficacy of immunomodulatory treatments including fingolimod (Fujino et al., 2003), natalizumab (Yednock et al., 1992), and glatiramer (Teitelbaum et al., 1971).

Fingolimod is a sphingosine 1-phosphate receptor agonist (Brinkmann et al., 2002) and shows immunosuppressive activity in animal models of graft rejection and autoimmune diseases (Chiba et al., 1998; Matsuura et al., 2000). It has been reported that administration of fingolimod starting from the day of EAE induction suppressed the development of an autoimmune response (Brinkmann et al., 2002; Fujino et al., 2003). The administration of fingolimod after EAE onset improved the clinical score of neurological symptoms (Webb et al., 2004). The effectiveness of fingolimod is ascribable to the inhibition of T-cell infiltration into the central nervous system (Kataoka et al., 2005). It has been also shown that fingolimod promotes the restoration of blood brain barrier via its effects on microvascular cells (Balatoni et al., 2007; Foster et al., 2009).

We have reported a three-dimensional analysis of a mouse brain hemisphere using synchrotron radiation microtomography (Mizutani, Saiga, Ohtsuka et al., 2016). Axonal networks in the brain hemisphere were reconstructed from three-dimensional images of lyophilized tissues. Since water is the primary component of soft tissue, its removal by lyophilization can recover the intrinsic X-ray contrast of myelin and enable visualization of the brain-wide network of axonal tracts in the mouse brain hemisphere. In this study, we used the same method to create a three-dimensional visualization of brain hemispheres and spinal cords of EAE-induced mice, which were raised with or without administration of fingolimod after the onset of the neurological symptoms. Microtomographic sections of spinal cords of the EAE mice showed dilation of capillary vessels and micrometer-scale vacuolation. Vacuolation was also observed in cerebellums of the EAE mice. Although a microtomographic study of the EAE mouse (Cedola et al., 2017) revealed thinning of the capillary wall, the relationship between the tissue structure and neurological symptoms have not been delineated. In order to analyze structural alterations in the spinal cord tissue, three-dimensional Cartesian coordinate models of vessel networks and vacuolation lesions were built by tracing the three-dimensional image according to the method used in the structural analysis of human brain neurons (Mizutani et al., 2018). The relationship between the tissue structure and clinical score was then quantitatively analyzed on the basis of the obtained Cartesian coordinate models.

## 2. Materials and Methods
### 2.1 EAE mouse

All mice were kept in the SPF animal facility at Tokai University School of Medicine and fed ad libitum under a 12:12 light and dark cycle. C57BL/6J mice nine weeks of age were purchased from CLEA Japan. The animal experiment of this study was approved by the Institutional Animal Care and Use Committee at Tokai University (approval number 161073) and performed in accordance with the institutional guidelines.

After acclimating the mice for eight days in the animal facility, EAE was induced using a Hooke Kit (Hooke Laboratories, MA, USA), as reported previously (Mendel et al., 1995). First, ten mice were immunized subcutaneously with an emulsion of $MOG_{35-55}$ in complete Freund's adjuvant, and followed by intraperitoneal administration of pertussis toxin dissolved in phosphate-buffered saline. Since one of the mice showed a leakage of the injected emulsion, it was excluded from the experiment. The first injection day was referred to as day 0. On day 1, pertussis toxin was administered again intraperitoneally. Five control mice received injections of saline.

Mice were weighed daily and clinically scored on a scale of 0–5: 0 = no symptoms, 1 = limp tail, 2 = weakness of hind limbs, 3 = complete paralysis of hind limbs, 4 = complete paralysis of hind limbs and partial paralysis of forelimbs, and 5 = death or euthanized. In-between scores were recorded if the symptoms were between these definitions. Wet chow and hydrogel were placed on the cage base when the mice appeared to have difficulty feeding. The immunized mice showed paralysis symptoms from day 9 and reached an overall mean score of 2.9 on day 18. On that day, the immunized mice were divided into two groups in a balanced manner in accordance with their clinical scores. From the grouping day, one group consisting of five mice (F1–F5) received daily administration of fingolimod (Wako Chemical, Japan) saline at 1 mg/kg/day perorally. The other EAE group consisting of four mice (E1–E4) and the control group consisting of five mice (C1–C5) received saline. Mouse E1 of the non-treated group developed severe paralysis and was euthanized on day 25 with a subcutaneous injection of pentobarbital. The other mice were euthanized on day 28.

**2.2 Preparation of brain and spinal cord samples**

The brains of the EAE and control mice were dissected immediately after euthanasia under a perfusion fixation using formaldehyde saline. The right hemispheres of the dissected brains were embedded in paraffin to prepare Klüver-Barrera sections under the standard procedure. The left hemispheres were used in the X-ray microtomographic analysis. Vertebral segments of Th10–L3 corresponding to spinal cord segments of Th13–Co1 were dissected and immersed in formaldehyde saline. Then, the spinal cords were excised from the vertebra. The anterior and posterior ends of the excised spinal cords approximately corresponding to Th13 and S3–S4 segments were used for preparing Klüver-Barrera sections. The spinal cords of remaining segments were used in the X-ray microtomographic analysis.

The brain and spinal cord samples for the microtomographic analysis were dehydrated in an ethanol series and further soaked in t-butyl alcohol, as reported previously (Mizutani, Saiga, Ohtsuka et al., 2016). The samples were frozen at -20ºC overnight and then lyophilized for 1 hour using a JFD-310 freeze drying device (JEOL, Japan). The lyophilized tissue samples were attached to mounting rods by using epoxy glue.

**2.3 Microtomography**

Synchrotron radiation microtomography was performed at the BL20B2 beamline (Goto et al., 2001) of SPring-8. The samples were mounted on the rotation stage by using brass fittings. Transmission images were recorded with a CMOS-based X-ray imaging detector (AA40P and ORCA-Flash4.0, Hamamatsu Photonics, Japan) using a P43 scintillator screen. The field of view and effective pixel size of the detector were 5.63 mm × 5.63 mm and 2.75 μm × 2.75 μm, respectively. Transmission images of the spinal cord samples were acquired with a rotation step of 0.1° and exposure time of 500 ms using 8-keV monochromatic X-rays. The X-ray energy was calibrated with nickel foil. Since the sample length was longer than the viewing field height, multiple datasets were collected by shifting the sample along its longitudinal axis. Datasets of the brain hemisphere samples were collected with the offset CT setup, in which the sample rotation axis was placed at the right end of the viewing field and transmission images were acquired by rotating the sample through 360° and using 12-keV monochromatic X-rays. The rotation step per frame was 0.05°, and the exposure time per frame was 200 ms. The X-ray energy was calibrated with gold foil.

The obtained datasets were subjected to tomographic reconstruction using the RecView software available from https://mizutanilab.github.io/ under the BSD 2-Clause License. Microtomographic sections of multiple datasets were aligned and stacked in order to reconstruct a three-dimensional image of the entire sample. The spatial resolutions of the microtomographic sections were estimated to be 6–8 μm by using three-dimensional square-wave patterns (Mizutani et al., 2008) and Fourier domain plot (Mizutani, Saiga, Takekoshi et al., 2016). The mean X-ray linear attenuation coefficient (LAC) of the whole brain hemisphere was calculated from the volume showing LACs larger than 0.5 cm$^{-1}$. The mean X-ray LAC of the cerebellum was calculated by defining the cerebellum region with a polygon lasso.

**2.4 Structural analysis**

Since the microtomographic sections of the EAE and control mice showed differences in vacuolation and in the capillary vessel networks of the spinal cords, we built Cartesian coordinate models of the vacuolation lesions and vessel networks by using the method reported for geometric analysis of human brain neurons (Mizutani et al., 2018). The role of the data manager was allotted to RS, while the role of the data analyst was allotted to RM in order to eliminate human biases from the model building. The data manager reconstructed three-dimensional images from the microtomographic datasets and coded dataset names, but did not have access to the analysis results. The data analyst built Cartesian coordinate models of tissue structures from the three-dimensional images, but did not have access to the sample name. After the analyst finished building models for all of the samples, the Cartesian coordinate files were locked down. The manager then disclosed the sample name to the analyst so that the tissue structures between the animal groups could be compared.

The models were built in five steps: 1) generation of the initial model of capillary vessel network, 2) manual editing of the working model, 3) model refinement with the conjugate gradient minimization, 4) automatic vacuole scanning, and 5) final examination. Steps 1) – 3) were performed as reported previously (Mizutani et al., 2018). In the generation of the initial model, the entire three-dimensional image was divided into image sections of 1.4 mm thickness, and each image section was scanned by calculating the gradient vector flow to find luminal structures. The luminal structures found in the scan were traced using a three-dimensional Sobel filter to build capillary vessel models. An attenuation coefficient threshold of -1.5 cm$^{-1}$ was used in the tracing process. The computer-generated model was examined manually and edited to connect adjacent vessel structures, to remove models built into non-luminal structures, and to add vessel models that were not traced in the automatic model generation. The edited model was refined through conjugate gradient minimization of a target function that represents the fitness of the model to the image (Mizutani et al., 2018). Next, vacuolation spheres or ellipsoids with intensities lower than 0.5 cm$^{-1}$ and diameters of 20–40 μm were searched. The vessel models were used for masking the vacuolation search so as to avoid building vacuolar models in the vessel structure. Finally, the working model was examined manually to remove structures built outside the spinal cord.

A high-resolution MRI study of the vertebra of a C57BL/6J mouse has been reported (Harrison et al., 2013). A sagittal section of the vertebra indicated that the length of spinal cord segments of L1-L5 was 7.2–7.3 mm and the dorso-ventral thickness of the L1 segment was 1.5–1.6 mm. In the microtomographic section of the lyophilized spinal cord, the mean dorso-ventral thickness of the L1 segment was 0.996 mm, indicating that 64% shrinkage occurred in the sample preparation process. This estimation suggests that the length of L1–L5 segments in the lyophilized spinal cord is approximately 4.6 mm. Therefore, we analyzed structures within 4.6 mm from the L1 end. The geometry of the vessel structures was analyzed using the method for geometric analysis of human brain neurons (Mizutani et al., 2018). The mean diameter of the capillary vessels was calculated by averaging the diameters of the nodes of the vessel models.

Coordinate errors of protein crystal structures have been estimated from Luzzati plots (Luzzati, 1952) or sigma-A plots (Read, 1986). The coordinate errors of most crystal structures have been reported to be approximately 1/10th of their spatial resolution. In this study, we built and refined Cartesian coordinate models of tissue structures according to the methods of protein crystallography (Kleywegt & Jones, 1997). Although a method for evaluating the coordinate errors of tissue structures has not been established, we suggest that those of the obtained models were 0.6–0.8 μm, because the resolutions of the obtained images were 6–8 μm and the models were built according to the methods used in protein crystallography.

## 2.5 Statistical tests

Statistical tests of the clinical score and structural parameters were performed using the R software. Significance was defined as $p < 0.05$. Two-sided Welch's t-tests were used to examine the equality of the clinical score and structural parameters between animal groups. Pearson's correlation coefficients were calculated for analyzing correlations between the structural parameters and the clinical score.

## 3. Results
### 3.1 EAE induction

The clinical scores of each mouse are summarized in Table 1. The EAE-induced mice showed symptoms of neurological disorder, such as paralysis of limbs and disability of postural maintenance. The control mice showed no neurological symptoms. The mean clinical score of the final three days of the fingolimod-treated EAE mice (1.9 for 5 mice) was lower than that of the non-treated mice (3.0 for 4 mice), although the difference was insignificant (Welch's t-test, $p = 0.23$).

**Table 1.** Clinical score and structural parameters of EAE-induced and control mice

| Group | Mouse ID | Mean clinical score | | X-ray attenuation coefficient ($cm^{-1}$) | | Spinal cord structure | | |
|---|---|---|---|---|---|---|---|---|
| | | From onset | Final 3 d | Whole brain | Cere-bellum | Number of vacuoles | Number of vessel nodes | Vessel diameter (μm) |
| EAE / non-treated | E1 | 2.9 | 4.2 | 1.47 | 1.64 | 1221 | 30886 | 9.5 |
| | E2 | 2.8 | 3.5 | 1.49 | 1.67 | 2241 | 26846 | 9.4 |
| | E3 | 2.3 | 3.0 | 1.50 | 1.67 | 2917 | 32299 | 9.7 |
| | E4 | 1.5 | 1.2 | 1.49 | 1.61 | 664 | 35776 | 8.8 |
| EAE / fingolimod | F1 | 2.7 | 2.5 | 1.51 | 1.68 | 3631 | 24839 | 9.2 |
| | F2 | 0.8 | 1.2 | 1.49 | 1.65 | 729 | 27753 | 8.7 |
| | F3 | 2.2 | 2.2 | 1.48 | 1.62 | 2043 | 25940 | 8.8 |
| | F4 | 2.9 | 3.0 | 1.42 | 1.58 | 2092 | 23247 | 9.1 |
| | F5 | 1.0 | 0.8 | 1.52 | 1.68 | 742 | 34947 | 8.4 |
| Control | C1 | 0.0 | 0.0 | 1.41 | 1.62 | 336 | 4902 | 8.5 |
| | C2 | 0.0 | 0.0 | 1.52 | 1.65 | 392 | 27654 | 8.3 |
| | C3 | 0.0 | 0.0 | 1.48 | 1.64 | 282 | 19268 | 8.4 |
| | C4 | 0.0 | 0.0 | 1.45 | 1.65 | 357 | 37952 | 7.7 |
| | C5 | 0.0 | 0.0 | 1.46 | 1.67 | 293 | 19396 | 7.8 |

### 3.2 Microtomography and histology of brain hemisphere and spinal cord

Figure 1 shows microtomographic cross sections of the L1 spinal cord segments of EAE-induced and control mice. Histologies of adjacent tissue sections are also shown. The major differences between the cross sections of the EAE and control mice are the dilation of capillary vessels and micrometer-scale vacuolation showing low X-ray attenuation coefficients. A microtomographic section of the spinal cord of a non-treated EAE mouse with severe paralytic symptoms (Figure 1a) showed vacuolated and damaged tissue structures mainly in the ventral white matter of the spinal cord. The vacuolation lesions were also discernible in the Klüver-Barrera section (Figure 1b). Myelin vacuolation or vesiculation is the common pathologic alteration of the myelin sheath in demyelinating diseases (Duncan & Radcliff, 2016). Microtomographic and histological sections of a control mouse (Figure 1e and 1f) showed a homogeneous and dense structure without vacuolation. A fingolimod-treated EAE mouse with moderate paralytic symptoms (mouse F5) showed fewer vacuolation lesions (Figure 1c and 1d) compared with a non-treated EAE mouse with severe symptoms (mouse E2; Figure 1a and 1b). The clinical score of this fingolimod-treated F5 mouse (total mean of 1.0 and final 3-day mean of 0.8; Table 1) suggested that the autoimmune response and resultant structural degeneration in the spinal cord tissue were moderate.

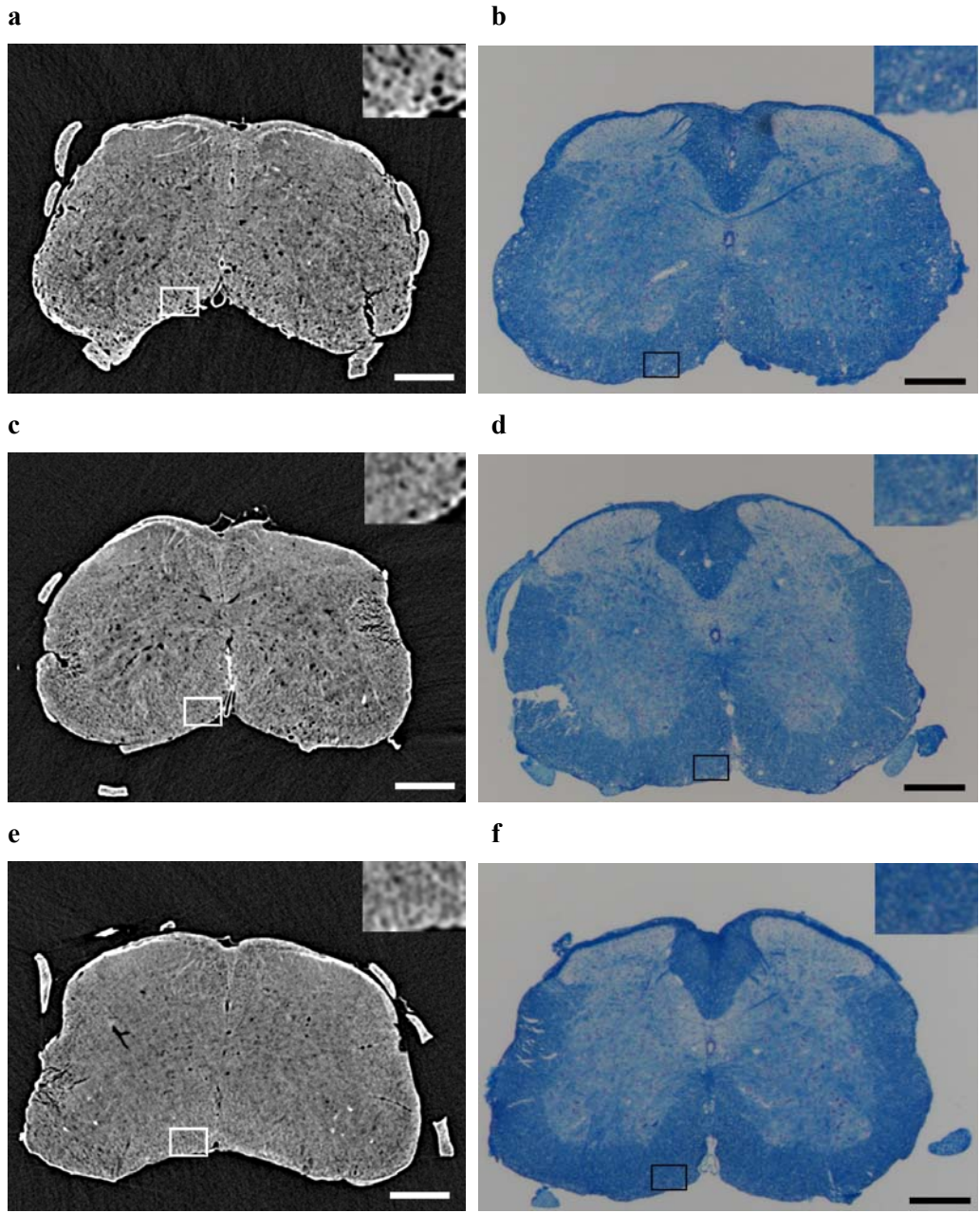

**Figure 1.** Microtomographic cross sections of L1 spinal cord segments and Klüver-Barrera sections of adjacent tissues. Linear attenuation coefficients from -0.5 cm$^{-1}$ to 3.5 cm$^{-1}$ were rendered with gray scale. Paraffin sections 5 μm in thickness were prepared for Klüver-Barrera staining. Scale bars: 250 μm. Insets in the upper right show three-fold magnification of areas indicated with boxes. (a) Microtomographic section of non-treated EAE mouse E2. (b) A Klüver-Barrera section of non-treated EAE mouse E2. (c) Microtomographic section of fingolimod-treated EAE mouse F5. (d) Klüver-Barrera section of fingolimod-treated EAE mouse F5. (e) Microtomographic section of control mouse C1. (f) Klüver-Barrera section of control mouse C5.

Figure 2 shows a three-dimensional rendering of the brain hemisphere and cross sections of the cerebrum and cerebellum. Micrometer-scale vacuolation was observed in the cerebellar white matter of the non-treated EAE mouse (Figure 2d). This indicates that the tissue degeneration observed in the spinal cord also progressed in the cerebellum. In contrast, cerebrum sections showed no apparent structural differences between the EAE mice and the controls (Figure 2b and 2c) and shared similar structures of axonal tracts. This suggested that the inflammatory response had not reached the cerebrum during the mouse raising period. X-ray attenuation coefficients of the whole brain hemisphere and the cerebellum are summarized in Table 1. The mean attenuation coefficients of the cerebellum were higher than those of the whole hemisphere, although no significant differences were found between the EAE and the control mice (Welch's t-test, $p = 0.34$ for the whole hemisphere). It has been reported that lipid localization in axonal tracts can be visualized as a contrast of the X-ray attenuation coefficient (Mizutani, Saiga, Ohtsuka et al., 2016). These results indicated that the total lipid content was retained in the brain tissue even after EAE induction.

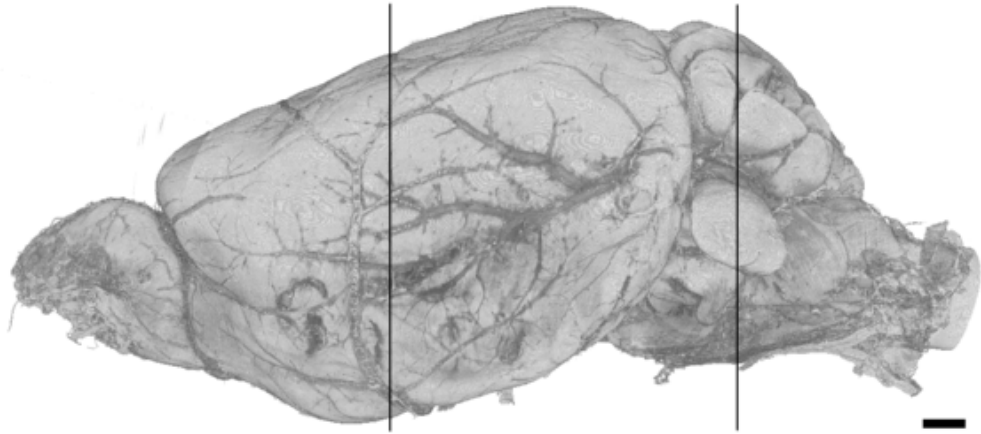

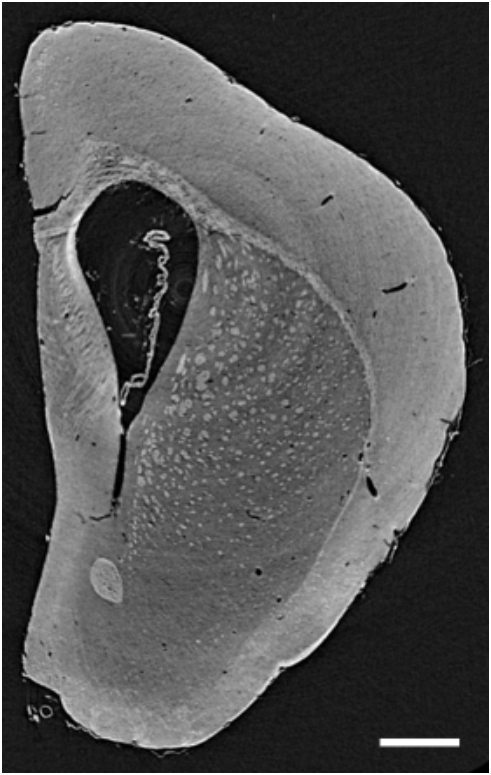
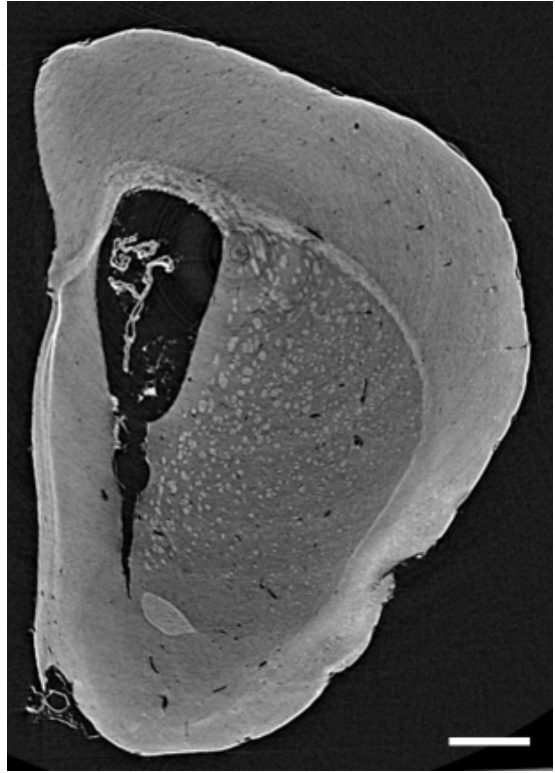

**Figure 2.** Structures of brain hemispheres of a non-treated EAE mouse and a control mouse. Scale bars: 500 μm. (a) Three-dimensional rendering of the left brain hemisphere of control mouse C5. Linear attenuation coefficients from 0.5 cm$^{-1}$ to 3.5 cm$^{-1}$ were rendered with gray scale. Approximate section positions are indicated with lines. (b) Coronal section of the cerebrum of non-treated EAE mouse E3. Linear attenuation coefficients from -0.5 cm$^{-1}$ to 3.5 cm$^{-1}$ were rendered with gray scale. (c) Coronal section of the cerebrum of control mouse C5.

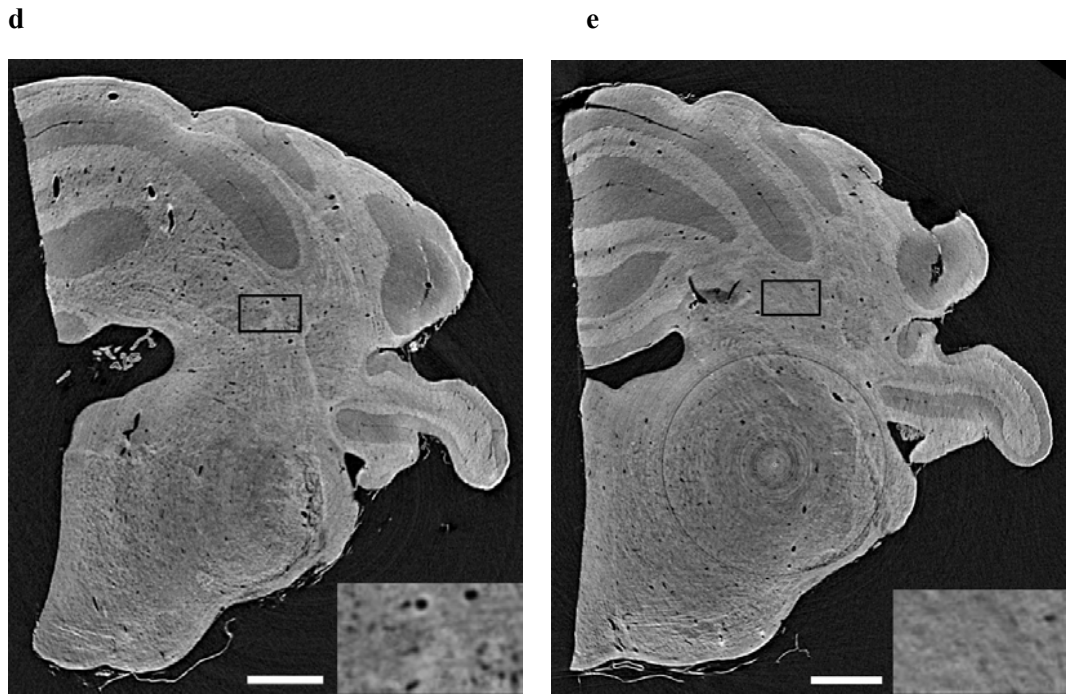

**Figure 2 (cont'd).** Structures of brain hemispheres of a non-treated EAE mouse and a control mouse. Scale bars: 500 μm. (d) Coronal section of the cerebellum of EAE mouse E3. (e) Coronal section of the cerebellum of control mouse C5. Insets on the lower right are three-fold magnifications of the cerebellar white matter indicated with boxes. Micrometer-scale vacuoles showing low X-ray attenuation coefficients were observed in the white matter.

**3.3 Analysis of three-dimensional tissue structures**

Since the microtomographic sections of the EAE mice showed structural alterations such as the vacuolation and the capillary dilation in the spinal cord tissue, we built Cartesian coordinate models of vessel networks and vacuolation lesions. The three-dimensional image of the spinal cord (Figure 3a) was used for building models of vessel network and vacuolation. The model building was performed according to the method used for analyzing neuronal structures of human cerebral tissues (Mizutani et al., 2018). The initial model of vessel network (Figure 3b) was generated using a three-dimensional Sobel filter and the gradient vector flow. The computer-generated model was superposed with the three-dimensional image and manually edited. The edited model was refined with conjugate gradient minimization (Figure 3c). The obtained vessel model was then subjected to vacuole scanning, in which vacuolation lesions with diameters of 20–40 μm were automatically built (Figure 3d). Since vessels and vacuoles were built by placing and connecting nodes in the three-dimensional coordinate space (Figure 3e), the geometry of the three-dimensional tissue structure could be analyzed quantitatively from the model coordinates.

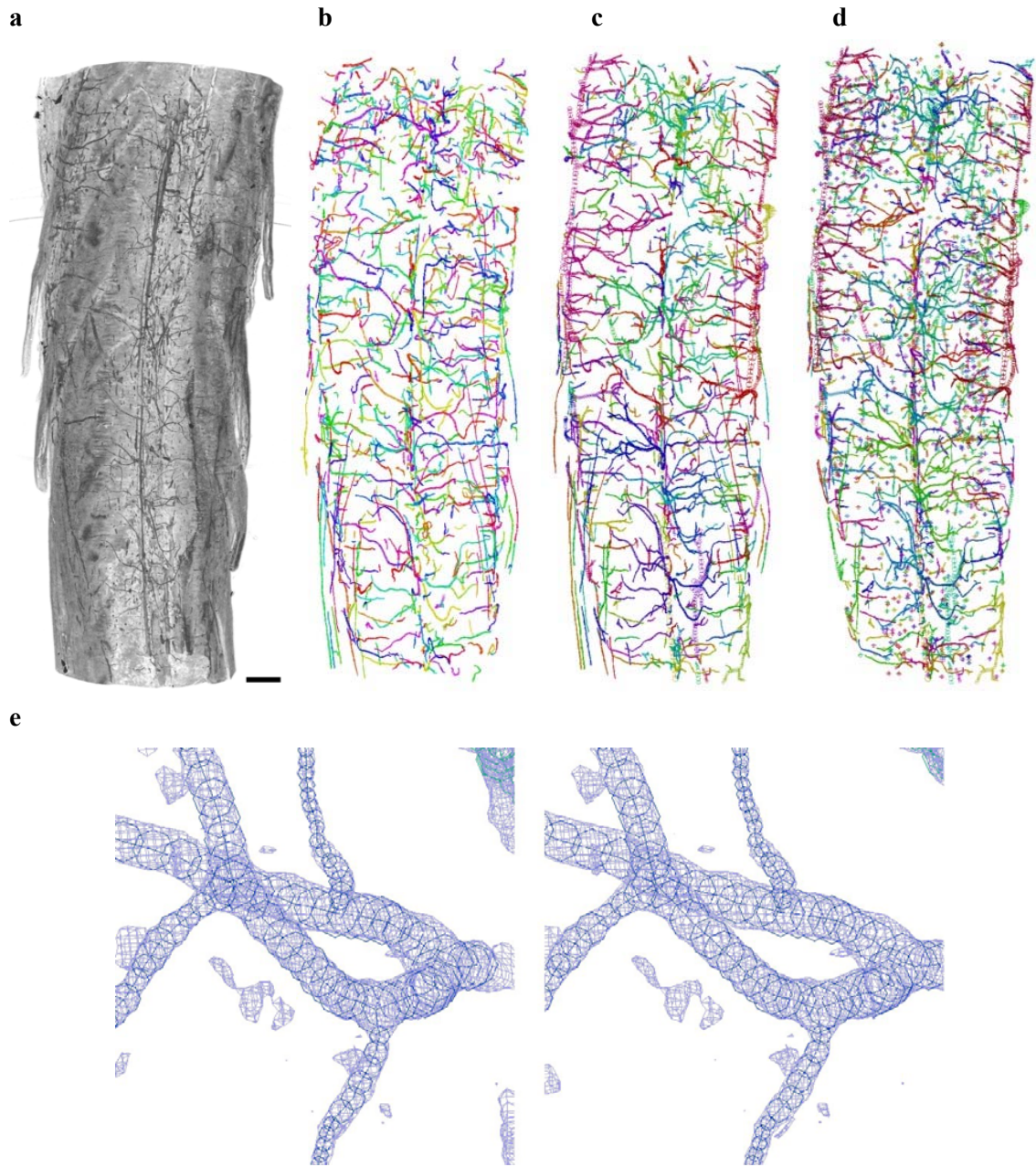

**Figure 3.** (a) Dorsal view of spinal cord of a non-treated EAE mouse (mouse E1). Linear attenuation coefficients from -3.5 cm$^{-1}$ (black) to 0.0 cm$^{-1}$ (white) were rendered with an inverse gray scale to visualize the internal vessel network. Scale bar: 250 μm. (b) The initial model was generated by automatically tracing luminal structures in the three-dimensional image. Structures are color-coded. (c) The computer-generated model was examined and edited manually to build the vessel network model. (d) Vacuolation lesions indicated with dots were identified by scanning the three-dimensional image. (e) Stereo drawing of a vessel model (blue) superposed on its three-dimensional map of linear attenuation coefficients (purple). The coefficient map was contoured at -1.0 cm$^{-1}$ with a grid size of 2.75 μm.

Figure 4 shows Cartesian coordinate models of vessels and vacuolation lesions in the spinal cords. These models illustrate that the EAE mouse had more vacuoles than the control had. Statistics of the tissue structures are summarized in Table 1. The number of vacuoles in the EAE mice was significantly higher than that of controls (Welch's t-test, $p = 0.0029$), though the amount of vacuolation did not exactly match the severity of the neurological symptoms. The major fraction of vacuoles was localized in specific regions of the spinal cord tissue (Supporting Information, Movie 1), presumably representing localization of neuroinflammation loci. The mean diameter of the capillary vessels was 9.4 μm for the non-treated EAE mice, 8.8 μm for the fingolimod-treated EAE mice, or 8.1 μm for the controls. The vessel diameters of the EAE mice were significantly higher than those of the controls (Welch's t-test, $p = 0.0019$). The difference in vessel diameter between the non-treated and fingolimod-treated EAE mice was not significant ($p = 0.066$), though the lack of significance is ascribable to the small size of the animal groups.

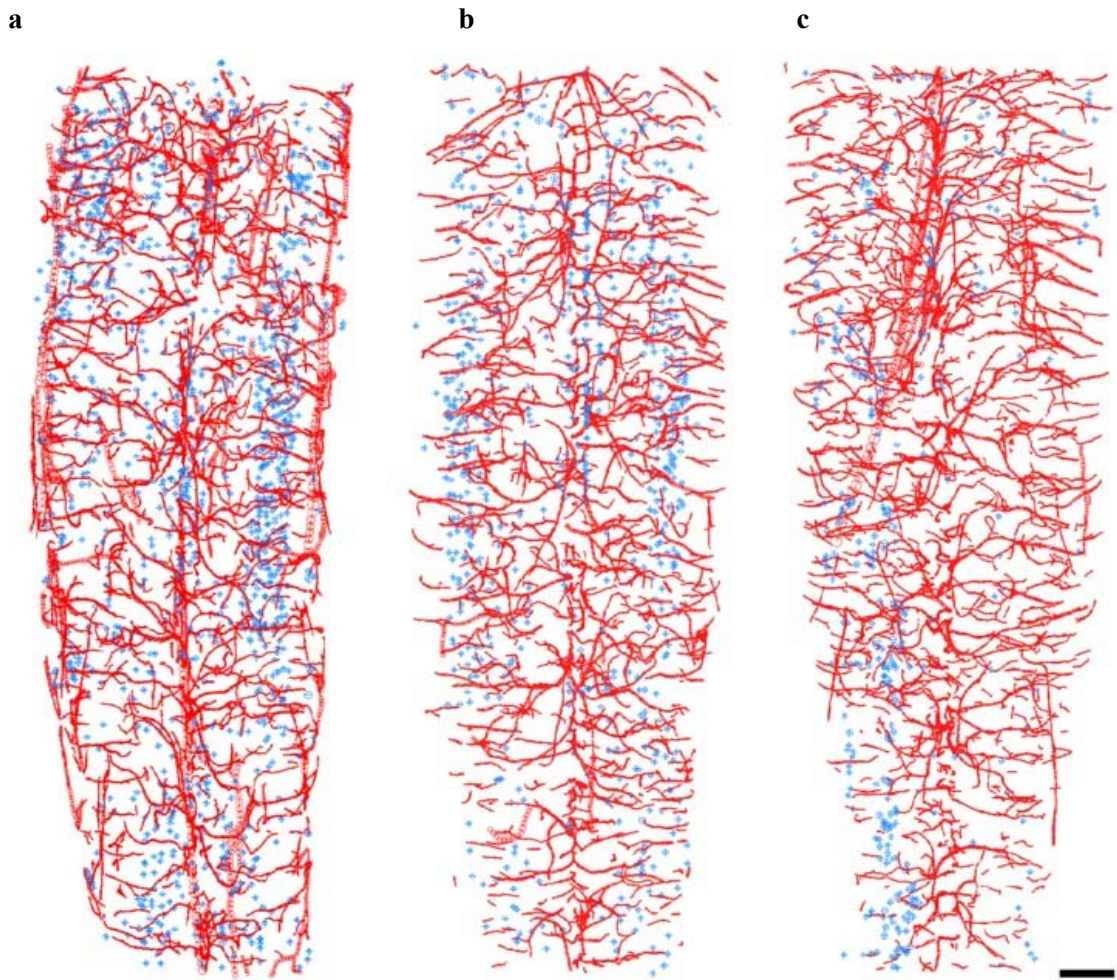

**Figure 4.** Vessel network and vacuolation of (a) non-treated EAE mouse E1, (b) fingolimod-treated EAE mouse F2, and (c) control mouse C4. Vessel networks are drawn in red and vacuolation lesions in blue. Scale bar: 250 μm.

A scatter plot of the vessel diameter versus the mean clinical score of the final three days is shown in Figure 5a. The plot illustrates a linear correlation between the vessel diameter and the clinical score (Pearson's correlation coefficient $r = 0.89$, $p = 1.6 \times 10^{-5}$). This indicates that the capillary dilation in the spinal cord is relevant to the neurological symptoms of the EAE mice. A plot of the number of vacuoles versus the clinical score is shown in Figure 5b. The number of vacuoles shows a stronger correlation with the mean score from onset to euthanasia (Pearson's correlation coefficient $r = 0.83$, $p = 0.00020$) than with the mean score of the final three days (Pearson $r = 0.75$, $p = 0.0019$). This suggests that vacuolation represents the cumulative effect of the inflammatory response. Other geometric parameters, including length, curvature, and torsion of the capillary vessel network, showed no remarkable correlation with the clinical score.

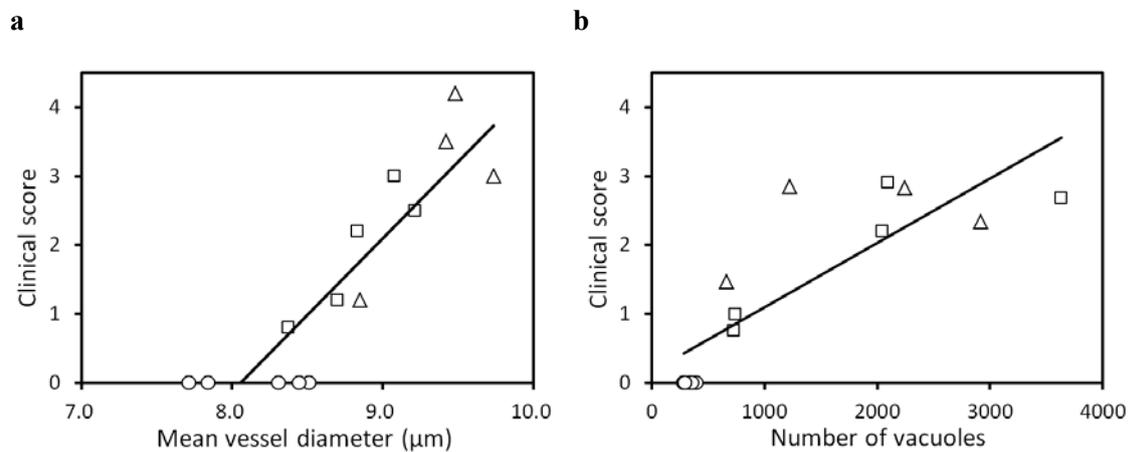

**Figure 5.** (a) Scatter plot of mean vessel diameter versus mean clinical score of the final three days before dissection. (b) Scatter plot of number of vacuoles versus mean clinical score from onset to euthanasia. Non-treated EAE mice are represented with triangles, fingolimod-treated EAE mice with squares, and controls with circles. Lines indicate linear regressions for all mice.

## 4. Discussion

The microtomographic analysis of the EAE mice indicated that 1) vacuolation lesions were generated in the spinal cord and the cerebellar white matter of the EAE mice, and that 2) the vessel diameter of the spinal cord showed a linear correlation with the clinical score of paralysis symptoms.

Myelin vacuolation is a common pathologic alteration of the myelin sheath (Duncan & Radcliff, 2016) and has been reported to be one of four ultrastructural patterns of myelin destruction in MS (Lucchinetti et al., 1996). A rat model of neuromyelitis optica, in which EAE was induced with an anti-aquaporin-4 antibody, indicated that vacuolation was generated in the spinal cord tissue (Kurosawa et al., 2015). A rabbit model of spinal cord ischemia showed a correlation between the amount of vacuolation and motor function score (Kurita et al., 2006). The results of this study revealed that the number of vacuolations was correlated with the mean clinical score from the onset of EAE to euthanasia. Therefore, the amount of vacuolation can be a measure of the total inflammatory activity over the disease course.

The linear correlation between the vessel diameter and clinical score indicates that vessel dilation is relevant to the neurological symptoms of the EAE mice. The vessel dilation in the EAE mice is ascribable to vasodilative mediators associated with inflammation or to congestion of the spinal cord tissue. It has been reported that the vessel network of the spinal cord of an EAE mouse appeared ramified and dense (Cedola et al., 2017). Some cardiovascular mediators, such as sphingosine-1-phosphate, promote lymphocyte egress and hence are likely to be responsible for the inflammatory response in the spinal cord tissue (Camm et al., 2014).

It has been suggested that the pathology of MS is the consequence of impaired venous drainage from the central nervous system (D'haeseleer et al., 2011). An MRI study indicated that the high inflammatory lesions in MS are associated with increased cerebral blood volume (Bester et al., 2015). Arterial transit time of cerebral blood flow was reported to be longer in MS patients with greater disability than in controls (Paling et al., 2014). These reports suggest that alterations in cerebral blood flow are a marker of MS pathology. It has been reported that nitric oxide synthase is expressed at significant levels in the brains of MS patients (Bagasra et al., 1995; Chuman, 2006). It has been also shown that an inhibitor of nitric oxide synthase can reduce the severity of EAE (Zhao et al., 1996). Therefore, vasodilative mediators, such as nitric oxide, should contribute to the pathology of MS. This is consistent with the results of this study indicating that neurological symptoms correlate with vasodilation.

The fingolimod used in this study causes persistent vasoconstriction in long-term treatment of MS (Camm et al., 2014), suggesting the relevance of the vasculature structure in MS. Although disease-modifying therapeutics help to control the symptoms of MS, they do not cure the disease or reverse the damage (Loma & Heyman, 2011). As most MS therapeutics have been developed so as to suppress the immune-mediated inflammation (Dendrou et al., 2015), other mechanisms of action should be explored in order to improve the treatment of MS. We suggest that vasodilation control may be an alternative path along which to pursue a cure of MS.

**Acknowledgments**

We thank Yoshiko Shinozaki (Support Center for Medical Research and Education, Tokai University) for assistance in raising and dissecting the mice. We thank Masayoshi Tokunaga (Support Center for Medical Research and Education, Tokai University) for assistance in preparing the tissue samples. We thank Tsutomu Hirose (Technical Service Coordination Office, Tokai University) for assistance in preparing adapters for microtomography. This work was supported in part by Grants-in-Aid for Scientific Research from the Japan Society for the Promotion of Science (nos. 21611009, 25282250, and 25610126). The synchrotron radiation experiments were performed at SPring-8 with the approval of the Japan Synchrotron Radiation Research Institute (JASRI) (proposal nos. 2017B1120 and 2018A1164).

**Author contributions**

R.M. and R.S. had full access to all the data in the study and take responsibility for the integrity of the data and the accuracy of the data analysis. R.M. designed the study according to suggestions from M.O. and S.T. regarding multiple sclerosis and its animal models. R.S., K.N., A.K., N.K., and R.M. prepared the tissue samples. R. S., M.H., A.T., K.U., and R.M. performed the synchrotron radiation experiment. R.S. managed the data analysis. R.M. analyzed the data and wrote the manuscript.

**Conflict of interests**

The authors have no conflicts of interest.

**ORCID**
Rino Saiga https://orcid.org/0000-0001-7427-7764
Shunya Takizawa https://orcid.org/0000-0002-4567-6384
Ryuta Mizutani https://orcid.org/0000-0002-5484-4861

**Supporting Information legends**

**Movie 1.** Three-dimensional structure of the vessel network and vacuolation distribution in the spinal cord of non-treated EAE mouse E1. Vacuolation lesions are indicated with crosses. Structures are color-coded.